\def\mn#1{\marginpar[]{#1}}
\def\comment#1{}

\documentstyle[pra,aps]{revtex}
\def\fsz{\footnotesize}
\title{Theory of
Brownian motion of Massive Particle in a Space with Curvature and Torsion\\
and Crystals with Defects}
\author{H. Kleinert and S.V. Shabanov\footnote{Alexander
von Humboldt fellow,
on leave from Laboratory
of Theoretical Physics, JINR, P.O.Box 79, Moscow, Russia
\\~\\
\sf Freie Universit\"at Berlin Preprint FUB--HEP/95--7\\
Email: kleinert@einstein.physik.fu-berlin.de;\,
shabanov@axp3.physik.fu-berlin.de\\
World Wide Web homepage: http://www.physik.fu-berlin.de/~kleinert
}
}
\address{Institut f\"{u}r Theoretische Physik,
        Freie Universit\"{a}t Berlin,
        Arnimallee 14, D-14195 Berlin}
\date{\today}

\begin{document}
\maketitle

\begin{abstract}
We develop a theory of
Brownian motion
of a massive particle,
including the effects of inertia (Kramers' problem),
in spaces with curvature and torsion.
This is done by invoking the recently
discovered generalized equivalence principle,
according to which the equations of motion
of a point particle in such spaces
can be obtained from the Newton equation in euclidean space
by means of a nonholonomic mapping.
By this principle,
the known
Langevin equation in euclidean space
goes over into the correct
Langevin equation in the Cartan space. This, in turn, serves to
 derive the Kubo and Fokker-Planck equations
satisfied by the particle distribution as a function of time in such a space.

The theory can be applied to classical diffusion processes
in crystals with defects.\\
\end{abstract}

{\bf 1}.
The classical equation of a massive point particle
in a thermal environment reads
\begin{equation}
m\ddot x_t^i=f_t^i+ \bar f_t ^i,
\label{}\end{equation}
where $f^i_t$ is an arbitrary time-dependent external force and
$ \bar f ^i_t$ a stochastic force caused by the thermal fluctuations
(we use subscripts for the time variable).
The stochastic force may be modeled by a bath of harmonic oscillators
of all frequencies $ \omega $ as
$
\bar{f}^i_t = \int_{0}^{\infty}d\omega\lambda_\omega
\dot{X}^i_{\omega t}\label{tforce}.
$
The oscillator coordinates
satisfy the equations of motion
$
\ddot{X}^i_{\omega t} + \omega^2X^i_{\omega t}=\lambda_\omega \dot x^i_t
$,
the right-hand side arising from the back-reaction of the particle.
Solving the latter equation with respect to $X^i_{\omega t}$ we find
$\bar{f}_t^i$ as a functional of $\dot x^i_t$.
Assuming an equal coupling of  all oscillators,
$ \lambda _ \omega \equiv  \sqrt{ 2 \gamma /\pi}$,
we obtain the stochastic differential equation
\begin{equation}
m\ddot{x}^i_t
+\gamma \dot{x}^i_t -f^i_t = \eta^i_t \ ,
\label{leeu}
\end{equation}
where $\eta_t^i$ is called the {\em noise variable\/}.
This has the
decomposition
\begin{equation}
\eta^i_t = \int\limits_{0}^{\infty}
d\omega \lambda_\omega
\left[\dot{X}^i_\omega\cos\omega t -X^i_\omega\sin \omega t\right].
\label{noise}
\end{equation}
For any given
phase space distribution
$\rho^B =\rho^B(\dot{X}, X)$
of initial oscillator velocities
$\dot{X}^i_\omega$ and coordinates $X^i_\omega$,
Eq.(\ref{leeu}) becomes
a
classical Langevin
equation with noise averages being
defined as mean-values with
respect to the distribution $\rho^B$. In thermal equilibrium,
$\rho^B$ follows the Boltzmann law
$\sim \exp(-H^B/kT)$, where
 the bath Hamiltonian is a sum of the oscillator energies
 $H^B =\int d\omega
  (\dot{X}_\omega^2 +\omega^2X^2_\omega)/2$. In this case,
the noise is  Gaussian and completely specified by its
vanishing expectation and its two-point correlation function:
\begin{equation}
\langle\eta_t^i\rangle =0\ ,\ \ \ \ \langle\eta_t^i\eta_{t'}^j\rangle =
6\gamma kT \delta^{ij}\delta(t - t')\ .
\label{cor}
\end{equation}

{\bf 2}.
By a nonholonomic mapping  $dx ^i=e^i{}_ \mu(q) dq ^\mu $
\cite{nh}, \cite{schouten}
the euclidean space
is carried into
an arbitrary metric-affine space
with curvature and torsion \cite{cartan}.
The geometry in this space is defined by
the metric $g_{ \mu  \nu }=e^i{}_ \mu
e^i{}_ \nu $ and the affine connection
$ \Gamma _{ \mu \nu }{}^ \lambda=e_i{}^ \lambda \partial _ \mu
e^i{}_ \nu $  \cite{schouten}.
The torsion is carried by the antisymmetric part of the affine connection,
$S_{\mu\nu}^{~~\, \lambda }
= (\Gamma_{\mu\nu}^{~\,~ \lambda }- \Gamma_{\nu\mu}^{\,~~ \lambda })/2$,
which is a tensor,
in contrast to $\Gamma_{\mu\nu}^{~\,~ \lambda }$ itself.
The curvature tensor arises from noncommuting derivatives
of $e^i{}_ \lambda$:
$R_{ \mu  \nu  \lambda }{} ^ \kappa =e_i{}^ \kappa (\partial _ \mu \partial_
\nu  -\partial_ \mu  \partial _ \nu )e^i{}_ \lambda $.
Thus, in a curved space, the nonholonomic quantities
$e_i{}^\mu(q)$ fail to
satisfy Schwarz' criterion and are
 no proper functions \cite{nh}.

Recently it was found that
the nonholonomic mapping
 $dx ^i=e^i{}_ \mu(q) dq ^\mu $
transforms  euclidean equations of motion, Schroedinger equations,
and path integrals  correctly into spaces with curvature and torsion
\cite{PI}. This universal mapping rule was called {\em quantum equivalence
principle\/} (QEP) \cite{com}.

Applying this
principle
to the
equation of motion
(\ref{leeu}), we obtain the analogous equation
in the general  metric-affine space:
\begin{equation}
m\left(\ddot{q}^\mu_t +
\Gamma_{\sigma\nu}^{~~\,\mu}\dot{q}^\sigma_t\dot{q}^\nu_t\right)
+\gamma \dot{q}^\mu_t -f^\mu_t = e_i{}^\mu(q_t)\eta^i_t \ .
\label{le}
\end{equation}
To obtain physical consequences,
we must find equations
in which
the nonholonomic
quantities
$e_i{}^\mu(q_t)$ are eliminated in favor of the
well-defined geometrical objects,
the metric
$g_{ \mu  \nu } (q_t)$ and the affine connection
$ \Gamma _{ \mu \nu }{}^ \lambda(q_t)$.
This is possible by deriving from (\ref{le})
Kubo's
stochastic Liouville equation
and the Fokker-Planck equation.

For this purpose
we rewrite the Langevin equation as a system of
two first-order differential equations
\begin{eqnarray}
\dot{q}^\mu_t &= &\frac{1}{m}g^{\mu\nu}(q_t)p_\nu^t\ ; \label{h1}\\
\dot{p}_\mu^t &= &-\frac{1}{m}
\left(\Gamma^{\sigma\nu}_{~~\,\mu}- g^{\sigma\lambda}
g^{\nu\alpha}\partial_\lambda g_{\mu\alpha}\right)p_\sigma^t
p_\nu^t -\frac{\gamma}{m}p_\mu^t + f^\mu_t + e^\mu_i\eta^i_t\ \nonumber\\
&\equiv& -F_\mu^t + e_\mu^i\eta^i_t\  .
\label{h2}
\end{eqnarray}
The last equation defines the total apparent force $F^t_\mu $,
where $f^t_\mu =g_{\mu\nu}(q_t)f^\nu_t$.
We mark the time dependence also by a superscript
to avoid a pile-up of subscripts.
Note that although the time derivative of the momentum $\dot{p}_\mu^t $
and the force $F^t_\mu$ do not transform as vectors under
general coordinate transformations,
Eq. (\ref{h2}) is nevertheless covariant
under these transformations, just as
the Langevin equation (\ref{le}).

At each time $t$, the system following Eqs. (\ref{h1}) and
(\ref{h2}) is in a microscopic state with the distribution function
$\delta(p-p^t)\delta(q-q_t)$ (omitting the spatial indices $ \mu $).
This can be thought of as a conditional
distribution function determining the
distribution of $p$ and
$q$ for a given
initial distribution function
$\delta(p-p^0)\delta(q-q_0)$.
If the initial values of $p^t$ and $q _t$
 are distributed with the probability
density $\rho(p^0,q_0)$,  the distribution
function at any later time $t$ can be found by
an average over the
initial distribution
\begin{equation}
\rho_t^\eta(p,q)= \int dp^0dq_0\rho(p^0,q_0)\delta(p-p^t)\delta(q-q_t)\ ,
\label{dis}
\end{equation}
where $q_t=q_t(p^0,q_0)$ and $p^t= p^t(p^0,q_0)$ are
solutions of the system (\ref{h1}), (\ref{h2}) with the initial
conditions $p^{t=0} = p^0$ and $q_{t=0} = q_0 $.
Upon taking a time derivative of (\ref{dis}) and making use of both
(\ref{h1}) and (\ref{h2}) and of the identity
\begin{equation}
\frac{d}{dt}\delta(z-z_t)=\dot{z}_t\frac{\partial}{\partial z_t}
\delta(z-z_t)= -\frac{\partial}{\partial z}
\left[\dot{z}_t\delta(z-z_t)\right] =
 -\frac{\partial}{\partial z}
\left[V(z)\delta(z-z_t)\right] \ ,
\label{ide}
\end{equation}
where a dynamical variable $z_t$ (any of $p^t_\mu$ or $q^\mu_t$)
obeys the equation of motion $\dot{z}_t =V(z_t)$,
we find Kubo's stochastic Liouville
equation
\begin{equation}
\partial_t\rho^\eta_t = \hat{L}(\eta_t)\rho^\eta_t\ ,
\label{kubo}
\end{equation}
with $\hat{L}(\eta_t)$ being the noise-dependent Liouville operator
\begin{equation}
\hat{L}(\eta_t)= -\frac{\partial}{\partial q^\mu}\circ
g^{\mu\nu}(q)\frac{p_\nu}{m}
+\frac{\partial}{\partial p_\mu}\circ\left[F_\mu(p,q)
-e_\mu^i\eta^i_t \right]\ .
\label{lo}
\end{equation}
The symbol $\circ$ emphasizes the fact that the differential operators
 in front of it act on the product of the
function behind it with
$\rho^\eta_t$ in Eq. (\ref{kubo}).

It is straightforward to verify the invariance of the Liouville
operator and, hence, of Kubo's equation with respect to general
coordinate transformations.

A solution of Kubo's stochastic equation (\ref{kubo}) is a noise-dependent
distribution function which determines the
probability
to find a particle in an infinitesimal volume $dq$
by
\begin{equation}
dP_t(q)
=dq\int dp\langle\rho_t^\eta(p,q)\rangle \equiv dq\int dp\rho_t(p,q)\ .
\label{prob}
\end{equation}
It follows then from (\ref{dis}) that $\int dP_t(q) =\int dP_0(q) =1$,
i.e. the temporal evolution of the probability distribution described
by Kubo's equation (\ref{kubo}) is unitary.

The locality of the noise correlator (\ref{cor}) enables us
to derive
a Fokker-Planck equation
governing the temporal evolution
of the noise-averaged distribution $\rho_t(p,q)$.
For this
let us first calculate the average $\varphi_t^i =\langle\eta_t^i
\delta(p-p^t)\delta(q-q_t)\rangle$. Generalizing the
partial integration formula $\int e^{-a \eta ^2/2} \eta f( \eta )d\eta
=a^{-1}\int  e^{-a \eta ^2/2}f'( \eta )$
$=-a^{-1}\int  \partial _ \eta  e^{-a \eta ^2/2}f( \eta )$
to any Gaussian noise, we find
\begin{eqnarray}
\varphi_t^i &=& \int\limits^\infty_{-\infty}dt'\langle\eta_t^i\eta_{t'}^j
\rangle\left\langle\frac{\delta}{\delta\eta^j_{t'}}
\delta(p-p^t)\delta(q-q_t)\right\rangle  \nonumber\\
&=&-6\gamma kT \left\langle\left(
\frac{\partial}{\partial p_\mu}\circ
\frac{\delta p_\mu^t}{\delta\eta_t^i} +
       \frac{\partial}{\partial q^\mu}
\circ\frac{\delta q^\mu_t}{\delta\eta_t^i}\right)
\delta(p-p^t)\delta(q-q_t)\right\rangle\ , \label{id}
\end{eqnarray}
where we have used the explicit form of the two-point noise
correlation function (\ref{cor}) and the identity (\ref{ide}). To calculate the
variational
derivatives of dynamical variables with respect to noise in (\ref{id}),
we formally integrate (\ref{h1})  and (\ref{h2}):
\begin{eqnarray}
p^t_\mu &=&p_\mu^0 +
\int\limits_{0}^tdt'
\left(- F_\mu^{t'}
+e_\mu^i\eta^i_{t'} \right)\ ; \label{hi1}\\
q_t^\mu &=&q^\mu_0 +\frac 1m\int\limits_{0}^tdt'
g^{\mu\nu}(q_{t'})p_\nu^{t'}\ .
\label{hi2}
\end{eqnarray}
Taking the variational derivative of (\ref{hi1}), we obtain
\begin{equation}
\frac{\delta p_\mu^t}{\delta\eta_{t'}^i} =-
\int\limits_{t'}^tdt''\left[\frac{\delta F_\mu^{t''}}{\delta\eta^i_{t'}}
-\eta_{t''}^j \frac{\delta e^j_{\mu}(q_{t''})}{\delta\eta^i_{t'}}
\right]+
\int\limits_{t'}^tdt''e^\mu_i(q_{t''})\delta(t''-t') \ .
\label{vd1}
\end{equation}
The Langevin equation (\ref{h2}) is causal which implies that
$p^\mu_{t''}$ or $q^\mu_{t''}$ depend on $\eta^i_{t'}$ only
for $t''>t'$. This has been used to restrict the integration
range in (\ref{vd1}). The first integral in the left-hand side
of (\ref{vd1}) tends to zero as $t'$ approaches $t$, whereas
the second integral is equal to $e^\mu_i(q_{t'})\Theta(t-t')$ and, hence,
is not uniquely determined, since it involves
 the
value of Heaviside function at zero argument, $\Theta(0)$,
when $t=t'$. Therefore a
regularization
is needed. We replace the delta-function in  the correlator (\ref{cor})
by a smooth would-be delta-function of width $ \epsilon $,
 $\delta_\epsilon(t-t')$, which satisfies
$\int_{-\infty}^\infty dt\delta_\epsilon(t) =1$
and $\delta_\epsilon(t)=\delta_\epsilon(-t)$.
It is easily seen from (\ref{noise}) that such a regularization
can be achieved by retaining a weak dependence on
$ \omega $ in the  coupling constant $\lambda_\omega$
$\footnote{This physical regularization corresponds to the
Stratonovich treatment of stochastic integrals \cite{risken}, p.55.
A symmetrically smoothed correlator appears also due to quantum
effects (the fluctuation-dissipation theorem).}$.
With this regularization, we have
to replace $\delta p_\mu^t/\delta\eta_{t}^i $ in (\ref{id})
by
\begin{equation}
\int\limits_{-\infty}^\infty dt'\delta_\epsilon(t-t')
\frac{\delta p_\mu^t}{\delta\eta_{t'}^i}=
\int\limits_{-\infty}^\infty dt'\delta_\epsilon(t-t')
e_\mu^i(q_{t'})\Theta(t-t') =
\frac 12e_\mu^i(q_t)\ ,
\label{id2}
\end{equation}
where we have dropped the contribution of the first integral
of (\ref{vd1}) since it vanishes for $ \epsilon \rightarrow 0$.

Considering analogously
$\delta q^\mu_t/\delta\eta_{t'}^i $, we conclude that
the latter variational derivative vanishes as $t'$ approaches $t$.
Averaging  $\varphi^i_t$ with respect to the initial
distribution $\rho(p_0,q_0)$ we find the following relation
\begin{equation}
\int dp_0dq_0 \rho(p_0,q_0)\varphi^i_t =
\langle \eta_t^i\rho_t^\eta\rangle
= - 3\gamma kT e_\mu^i \frac{\partial}{\partial p_\mu}
\rho_t\ .
\label{id3}
\end{equation}
Taking the noise average of Kubo's stochastic equation (\ref{kubo})
and making use of (\ref{id3}), we end up with the Fokker-Planck
equation associated with the Langevin equation (\ref{le}):
\begin{eqnarray}
\partial_t\rho_t &=& \hat{L}_T\rho_t \ ,\label{fpe}\\
\hat{L}_T&=& -\frac{\partial}{\partial q^\mu}\circ g^{\mu\nu}
\frac{p_\nu}{m}+
\frac{\partial}{\partial p_\mu}\circ\left[ F_\mu(p,q)
+3\gamma kT g^{\mu\nu}\frac{\partial}{\partial p_\nu}
 \right]\ ,
\label{fplo}
\end{eqnarray}
Thus, we have eliminated the nonholonomic mapping
functions in favor of the well-defined affine connection and metric.
Integrating (\ref{fpe}) over all phase space we arrive at the
probability conservation law: $d/dt\int dpdq\rho_t =0$, where we
have used the fact of vanishing surface integrals occurring upon
the integration of the right-hand side of (\ref{fpe}).

In the overdamped limit, we
may drop the mass term in
(\ref{le}) and derive analogously an equation for the
$q$-space probability distribution $D_t(q)$,
defined by
$dP_t(q) = dq\sqrt{g}D_t(q)$,
\begin{eqnarray}
\partial_tD_t &=&\left[\frac{3kT}{\sqrt{g}}\partial_\mu
\circ e^\mu_i\partial_\nu\circ e^\nu_i\sqrt{g}
-\frac{1}{\gamma \sqrt{g}}\partial_\mu\circ f^\mu\sqrt{g} \right] D_t
\nonumber \\
&=&\left[\frac{3kT}{\sqrt{g}}\partial_\mu\circ g^{\mu\nu}
\sqrt{g}(\partial_\nu + 2S_\nu -\kappa f_\nu)\right] D_t\ ,
\label{damp2}
\end{eqnarray}
where $\kappa =(3kT\gamma)^{-1},\
S_\nu =S_{\nu\mu}^{~~~\mu}$ and $g= \det g_{\mu\nu}$. Since
$S_\nu$ and $f_\nu$ are vectors, the operator in the brackets
is invariant under general
coordinate transformations.
In the absence of torsion, the limiting equation (\ref{damp2})
agrees with
a result in Ref. \cite{zinn}.

{\bf 3}. Since defects in crystals
can be described geometrically
by a nonvanishing torsion and curvature in  a metric-affine space in which
distances are measured by counting atomic steps \cite{nh},
the Fokker-Planck equation
(\ref{fpe}) can be applied to the classical
diffusion of interstitial atoms in crystals \cite{HK}.

It must be realized, however,
that the approximation is necessarily very crude.
The reason for this is the twofold role played by the phonon bath.
First, it provides the system with a random driving force which
in our equations has been simulated by the model noise $\eta_t^i$.
Second, it influences the hopping probabilities of the interstitials.
The latter effect has not
been accounted for in our initial Eq. (\ref{le}).
In crystals, the hopping probability depends on elastic distortions
which correspond to time-independent general coordinate transformations
in the above theory. The Langevin equation (\ref{le}) is, however,
invariant with respect to these transformations.
A proper treatment should take into account the fact that
elastic distortions
are also produced by the off-shell phonons comprising the noise.

Another field of applications should be positrons or
 muons in a crystal at low temperatures
below a few or about 80 Kelvin, respectively.
These particles are bound in the lowest atomic orbits,
so that
their wave functions follow
the tight-binding approximation.
Spending most of
their time near the
atomic sites,
they live effectively in the above crystal geometry.

It is worth noting that in the absence of external forces, the overdamped
equation
(\ref{damp2})
agrees with an equation recently
studied
in Ref.~\cite{schmitz}.

{}~\\~\\
Acknowledgment:\\
The authors thank R. Schmitz
for sending them reprints of his work with
R. Bausch and L.A. Turski,
and
for subsequent discussions.
They also acknowledge a useful discussion with H. Grabert.

\end{document}